\documentclass[a4paper,fleqn,usenatbib]{mnras}

\usepackage[T1]{fontenc}
\usepackage{ae,aecompl}

\usepackage{graphicx}	% Including figure files
\usepackage{amsmath}	% Advanced maths commands
\usepackage{amssymb}	% Extra maths symbols
\usepackage{multicol}        % Multi-column entries in tables
\usepackage{bm}		% Bold maths symbols, including upright Greek
\usepackage{color}
\usepackage{xcolor}

\title[Maia variables]
{Maia variables and other anomalies among pulsating stars}
\author[L. A. Balona]{L. A. Balona\thanks{E-mail: lab@saao.ac.za}\\
\\
$^1$ South African Astronomical Observatory, 
P.O. Box 9, Observatory 7935, South Africa}

\begin{document}

\date{Accepted .... Received ...}

\pagerange{\pageref{firstpage}--\pageref{lastpage}} \pubyear{2011}

\maketitle

\label{firstpage}

\begin{abstract}
From {\em TESS} photometry, 493 mid- to late-B stars with high 
frequencies (Maia variables) have been identified.  The distribution of 
projected rotational velocities shows that the rotation rates of Maia variables 
are no different from those of SPB stars. Moreover, many Maia stars pulsate with 
frequencies exceeding 60\,d$^{-1}$. Rapid rotation is ruled out as a possible
factor in understanding the Maia variables.  There is clearly a serious problem
with current pulsational models.  Not only are the models unable to account for 
the Maia stars, they fail to account for the fact that SPB and $\gamma$~Dor 
variables form one continuous instability strip from the cool end of the $
\delta$~Sct region to the hot end of the $\beta$~Cep instability strip.  
Likewise, there is continuity between the distributions of $\delta$~Sct, Maia 
and $\beta$~Cep variables.  In fact, Maia stars seem to be an extension of
$\delta$~Sct stars to mid-B type. These observations suggest an interplay 
between multiple driving mechanisms rather than separate dominant mechanisms 
for each variability group.
\end{abstract}


\section{Introduction}

The Maia variables are defined as pulsating stars with high frequencies which 
are too hot to be classified as $\delta$~Scuti stars and too cool to be
$\beta$~Cephei variables.  These anomalous pulsating stars have been
suspected for many decades following a report by \citet{Struve1955} of 
short-period variations in the star Maia, a member of the Pleiades cluster. 
\citet{Struve1957} later disclaimed the variability. It is now known from
the {\it K2} space mission that Maia itself is a rotational variable with a 
10-d period \citep{White2017} and no sign of high frequencies.  Recently,
however, \citet{Monier2021c} has reported rapid light variations in the far 
ultraviolet.

In the past, ground-based observations by \citet{McNamara1985}, 
\citet{Lehmann1995}, \citet{Percy2000} and \citet{Kallinger2004} indicated
that Maia variables might exist, even though models do not predict pulsational 
instability.  From {\it CoRoT} observations, \citet{Degroote2009b} found 
several low-amplitude B-type pulsators between the SPB and $\delta$~Sct 
instability strips, with a very broad range of frequencies and low amplitudes.  

\citet{Mowlavi2013} provided further evidence for Maia variables.  They
found that many rapidly rotating mid-B stars in the open cluster NGC~3766 
pulsate with frequencies as high as 10\,d$^{-1}$. \citet{Mowlavi2016} also
found that the majority of these stars obey a period--luminosity relation.
They suggested the name ``FaRPB'' for these stars.  However, their properties 
are similar to what have been historically called the Maia variables.

Several B stars qualifying as Maia variables have been detected in the
{\it Kepler} field \citep{Balona2015c, Balona2016c}. Using {\em TESS} data,
\citet{Balona2020a} identified 131 Maia candidates.  More recently 
\citet{Gaia2022} found a population of Maia stars using {\em Gaia}
photometry. They interpret the high frequencies as an effect of rapid
rotation in SPB stars.

From {\em TESS} observations, \citet{Balona2020a} found that stars pulsating 
at frequencies higher than about 5\,d$^{-1}$ are common among all B stars, 
including late-B and early A stars.  Furthermore, there are no distinct 
regions of instability as expected from the models.  The $\beta$~Cep variables 
merge smoothly with Maia stars which merge smoothly into $\delta$~Sct stars.  
The same occurs for the low-frequency pulsators.  SPB stars are found all 
along the B-type main sequence and continue into the A star region where they 
merge with the hot $\gamma$~Dor variables. As a result, rather arbitrary 
effective temperature and frequency limits had to be introduced in order to 
define the different variability groups.

Using {\em TESS} photometry, \citet{Sharma2022} confirmed the continuity of 
SPB pulsations to cooler temperatures extending into the A stars.  They
suggested that these might be rapidly rotating SPB stars.  Rotation and 
uncertainties in effective temperature obviously contribute to the blurring of 
distinct instability regions, but it is still surprising that the concentration
of stars in their respective domains of instability expected from the models
is not seen, despite the large numbers of early-type pulsating stars 
discovered by {\em TESS}.  This is evident from Fig\,2 of \citet{Balona2020a}. 
  
When observed equator-on, rapid rotation lowers the apparent effective 
temperature due to equatorial gravity darkening, shifting the star to cooler
temperatures in the H--R diagram.  A rapidly-rotating $\beta$~Cep star could, 
for example, be mistaken for a Maia variable.  Furthermore,  gravito-inertial 
modes in moderate to fast rotators may have frequencies higher than normal. 
Thus a rapidly-rotating SPB star may also be mistaken for a Maia variable.
\citet{Salmon2014} investigated these effects and concluded that their
models could reproduce the observations of \citet{Mowlavi2013, Mowlavi2016}.

\citet{Saio2017}  examined pulsation models of rapidly-rotating main
sequence B stars and calculated the properties of prograde sectoral g and 
retrograde r modes excited by the $\kappa$~mechanism at the Fe opacity peak. 
They found that the period-luminosity relation described by
\citet{Mowlavi2016} can be explained by prograde sectoral g modes of rapidly 
rotating stars.

\citet{Daszynska-Daszkiewicz2017b} considered three hypothesis for the Maia
variables: rapidly rotating stars with underestimated masses, rapidly rotating 
stars with non-standard opacities and slowly rotating stars with non-standard
opacities.  While there are indications that one or more of these hypotheses
might be able to explain the observations, no definite conclusions can be
made at this stage.

It is clear that the most common opinion is that Maia variables are simply
rapidly rotating SPB stars.  It is therefore important to test this idea
by comparing the rotation rate of Maia stars with the rotation rate of SPB stars 
using the respective distributions of projected rotational velocities.  Such a 
test was performed by \citet{Balona2020a} using 41 Maia stars, but no 
significant difference in rotation rate between Maia stars and main sequence
stars could be found.  However, the sample of Maia stars is rather small 
for a definitive conclusion.

In this paper, we use {\em TESS} data from Sectors 1--59 to identify new
Maia candidates as well as other pulsating A and B stars.  Full details of
how the search was conducted, how estimates of effective temperature were
obtained and the sources of projected rotational velocities are described in 
\citet{Balona2022c}. The larger numbers of variable stars also allow a
better examination of the boundaries between various variability groups which
have been historically defined by their different pulsation mechanisms.  It
is show that the concept of distinct instability strips is not valid and
that the blending of these groups cannot simply be attributed to rapid
rotation or uncertainties in effective temperature.

\section{New Maia variables}

The {\it TESS} mission has been observing the whole sky and obtaining light 
curves for thousands of stars with two-minute cadence. This wide-band 
photometry has been corrected for long-term drifts using pre-search data 
conditioning  (PDC, \citealt{Jenkins2010b}).  The author has been engaged
in a project to classify the variability of stars hotter than about 6000\,K.
Results of the classification of over 120000 stars, together with their
effective temperatures, $T_{\rm eff}$, luminosities, $\log L/L_\odot$, and 
projected rotational velocities, $v\sin i$, are reported in 
\citet{Balona2022c}.  The values of $\log L/L_\odot$, are from Gaia DR3
parallaxes \citep{Gaia2016, Gaia2018}.

During this process, many stars were detected which normally would be
classified as $\delta$~Sct variables, but with effective temperatures in the
range 10000--18000\,K.  These were classified as Maia variables if frequency 
peaks higher than $\nu_{\rm min} = 5$\,d$^{-1}$ were detected.  The choices 
of $T_{\rm eff}$ range and $\nu_{\rm min}$ are arbitrary, but are guided by 
the fact that most SPB stars have frequencies not much higher than about 
3\,d$^{-1}$.  The full list of 493 Maia stars (and other variables) is to be 
found in \citet{Balona2022c}.  In the present work, only stars within the
main-sequence band are considered.  This band is defined by $-0.5 < 
\Delta\log{L/L_\odot} < 1.5$, where $\Delta\log{L/L_\odot}$ is the luminosity 
above the zero-age main sequence.  Out of the 493 Maia variables, 420 stars
are located within this main sequence band.

\begin{figure}
\begin{center}
\includegraphics{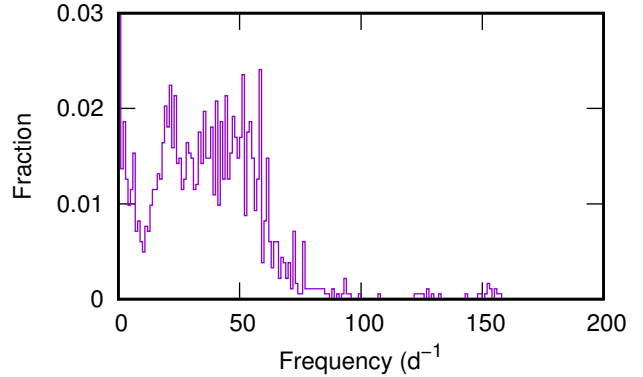}
\caption{Frequency distribution in Maia stars.}
\label{fdis}
\end{center}
\end{figure}

The choice of $\nu_{\rm min}$ is not too important.  With
$\nu_{\rm min} =  5$\,d$^{-1}$, 420 stars are classified as Maia variables.
If $\nu_{\rm min}= 10$\,d$^{-1}$ is used, then 351 stars are Maia variables,
but then one needs to modify the definition of SPB stars to include this new
upper frequency limit.  Clearly, most Maia variables have frequencies higher 
than 10\,d$^{-1}$.  

On the other hand, the boundary between SPB and $\beta$~Cep
stars is not at all well defined. Some bright, well-known $\beta$~Cep stars 
have frequencies lower than 4\,d$^{-1}$, in which case these would be
considered as SPB variables according to the above definition with
$\nu_{\rm min} = 5$\,d$^{-1}$.  To allow for this, the frequency limit for 
SPB stars with $T_{\rm eff} \ge 18000$\,K was lowered to $\nu_{\rm min} =
3$\,d$^{-1}$.  Thus in \citet{Balona2020a} and \citet{Balona2022c}, as well as 
in this work, the definition for SPB variables depends on whether the effective
temperature is greater or lower than 18000\,K.  \citet{Sharma2022}
encountered the same difficulty, and set $\nu_{\rm min} = 2.4$\,d$^{-1}$ to 
distinguish between SPB and $\beta$~Cep variables. 

The actual frequency distribution in Maia variables is shown in
Fig.\,\ref{fdis}. This figure was constructed from frequency peaks with 
signal-to-noise ratio S/N $>$ 5.0 irrespective of amplitude.  Frequencies
between 20--50\,d$^{-1}$ are quite common.  It is clear that such high 
frequencies in mid-B stars are incompatible with current models.  In their 
models of rapidly rotating SPB stars, \citet{Salmon2014} do not find any 
instance of pulsations with frequencies higher than 10\,d$^{-1}$. This is 
already an indication that rapid rotation should be discounted as an 
explanation for the Maia variables.

There are quite a few Maia stars which pulsate at rather high frequencies.  In 
fact, they would probably be mistaken for roAp stars, except that they are of 
type B and non-peculiar.  The roAp stars themselves can no longer be 
considered a separate class \citep{Balona2022a}.  As described in 
\citet{Balona2022c}, the class MAIAH is given if the highest frequency peak 
lies in the range 50--60\,d$^{-1}$, while if it is higher than 60\,d$^{-1}$, 
the notation MAIAU is used.  This does not indicate a separate class, but is a 
convenient way of identifying Maia stars with these peculiar high frequencies.
A total of 90 high-frequency Maia variables are listed in Table\,\ref{maiau}.
This consists of 43 MAIAH and 47 MAIAU stars.

\begin{table*}
\caption{List of Maia variables with high frequencies.  These are classified as
MAIAH if the maximum frequency $\nu_{\rm max} < 60$\,d$^{-1}$ and as MAIAU if 
$\nu_{\rm max} > 60$\,d$^{-1}$.  The number of frequencies greater than 
50\,d$^{-1}$, $N$, is also shown as well as the effective temperature, 
$T_{\rm eff}$ (in K) and the luminosity, $\log\tfrac{L}{L_\odot}$.  The last 
column is the spectral type.}
\label{maiau}
\begin{center}
\resizebox{17cm}{!}{
\begin{tabular}{rlrrrrlrlrrrrl}
\hline
\multicolumn{1}{c}{TIC} & 
\multicolumn{1}{c}{Var Type} & 
\multicolumn{1}{c}{$\nu_{\rm max}$} & 
\multicolumn{1}{c}{$N$} & 
\multicolumn{1}{c}{$T_{\rm eff}$} & 
\multicolumn{1}{c}{$\log\tfrac{L}{L_\odot}$} & 
\multicolumn{1}{l}{Sp. Type} &
\multicolumn{1}{c}{TIC} & 
\multicolumn{1}{c}{Var Type} & 
\multicolumn{1}{c}{$\nu_{\rm max}$} & 
\multicolumn{1}{c}{$N$} & 
\multicolumn{1}{c}{$T_{\rm eff}$} & 
\multicolumn{1}{c}{$\log\tfrac{L}{L_\odot}$} & 
\multicolumn{1}{l}{Sp. Type}  \\
\hline
  10433563 &  MAIAH              &  58.24 &     1 &   11300 &   2.14 &  B9.5V          &   262566258 &  MAIAH              &  58.76 &     1 &   11300 &   1.79 &  B9             \\
  27804376 &  MAIAU              &  60.52 &     2 &   11300 &   1.48 &  B9/A1IV/V      &   263261751 &  MAIAH              &  53.59 &     1 &   12600 &   1.75 &  B8:            \\
  30965889 &  MAIAU              &  60.97 &     1 &   10174 &   1.14 &  A0             &   264540595 &  SPB+MAIAH          &  50.22 &     2 &   11164 &   1.87 &  B9.5V          \\
  38127832 &  MAIAH+SXARI        &  54.56 &     6 &   12500 &   2.68 &  B9pSi          &   282742134 &  MAIAH              &  53.26 &    11 &   11300 &   1.25 &  B9             \\
  52831545 &  MAIAU+ROT          &  53.29 &    17 &   10053 &   1.16 &  A2             &   284163505 &  MAIAH+ROT          &  53.31 &     2 &   10257 &   2.03 &  B9.5Vp         \\
  52844490 &  MAIAH              &  51.64 &     7 &   10047 &   1.14 &  A2             &   284473460 &  MAIAU              &  51.32 &     7 &   11300 &   1.61 &  B9             \\
  60245596 &  MAIAH              &  54.18 &     2 &   14000 &   2.16 &  B7/8IV         &   284935176 &  MAIAH              &  72.21 &     3 &   11300 &   1.32 &  B9             \\
  74214081 &  MAIAU              &  93.77 &     1 &   11273 &   1.35 &  A0V            &   286344698 &  MAIAU              &  72.73 &     1 &   10518 &   1.30 &  A2V            \\
  76138809 &  MAIAU              &  62.31 &     4 &   11245 &   2.38 &  B9IIIn         &   293290586 &  MAIAU+ROT          & 156.89 &     1 &   12600 &   2.89 &  B8/9IV/V       \\
  86703658 &  MAIAU              &  68.63 &     1 &   17500 &   2.96 &  B5V            &   299123331 &  MAIAH              &  52.29 &     1 &   10029 &   1.44 &  A0             \\
  92736909 &  MAIAU              &  62.11 &     2 &   10695 &   1.20 &  A2V            &   299486936 &  MAIAH              &  51.23 &     1 &   10740 &   1.12 &  A5             \\
  92780981 &  MAIAU              &  54.74 &     2 &   11300 &   1.40 &  B9             &   300010961 &  MAIAH+ROT          &  58.35 &     6 &   13363 &   2.29 &  B8III          \\
 105262398 &  MAIAH              &  51.47 &     2 &   10466 &   1.29 &  A0             &   300493372 &  MAIAU              &  61.20 &     3 &   11220 &   1.38 &  A0             \\
 105889061 &  MAIAU              &  60.58 &     3 &   11300 &   1.32 &  B9             &   301100741 &  MAIAH+SXARI        &  50.18 &     4 &   17500 &   2.54 &  hB5HeB8pSi     \\
 105896213 &  MAIAU              &  67.44 &     2 &   11300 &   1.37 &  B9             &   304570125 &  MAIAU              &  63.79 &    11 &   10523 &   1.16 &  A0             \\
 115119794 &  MAIAU+ROT          &  63.84 &     6 &   10267 &   2.35 &  A0III/V        &   308769611 &  MAIAU              &  55.22 &    38 &   10116 &   1.68 &  A0V            \\
 121893547 &  MAIAH              &  53.65 &     1 &   11300 &   1.61 &  B9             &   315207705 &  MAIAU              &  63.91 &     1 &   17500 &   2.89 &  B5/7V:n:       \\
 124494015 &  MAIAU              &  61.37 &    21 &   11300 &   1.32 &  B9.5V          &   332856650 &  MAIAH              &  51.96 &     1 &   11960 &   2.40 &  B8/9           \\
 125080827 &  MAIAH              &  52.94 &     1 &   11300 &   2.18 &  B9.5IVnn       &   340356526 &  MAIAU              &  58.50 &     4 &   12500 &   2.20 &  B8IV/V         \\
 125977802 &  MAIAU              &  56.70 &     2 &   10903 &   1.63 &  A0V            &   345553381 &  MAIAU              &  70.58 &     1 &   11300 &   2.19 &  B9V            \\
 132923245 &  MAIAU+FLARE        &  78.01 &     1 &   11300 &   1.44 &  B9             &   352297130 &  MAIAU+ROT          &  62.99 &     6 &   11300 &   1.52 &  B9             \\
 133696007 &  MAIAH+SXARI        &  50.28 &     6 &   10843 &   1.85 &  A0Si           &   355775097 &  MAIAU              &  51.10 &     3 &   10744 &   1.33 &  A3/5IV:        \\
 133702466 &  MAIAH              &  53.94 &     1 &   10644 &   1.68 &  B8/A0          &   362653719 &  MAIAH              &  52.64 &     1 &   11300 &   1.36 &  B9             \\
 134860590 &  MAIAU+ROT          &  61.33 &    10 &   10928 &   1.23 &  A2V            &   363917122 &  MAIAH              &  52.91 &     1 &   10407 &   1.45 &  A2/3V          \\
 134861413 &  MAIAU              &  51.10 &     7 &   11230 &   1.29 &  A3             &   372913430 &  MAIAH              &  56.72 &     1 &   11772 &   2.22 &  B8.5V          \\
 136179360 &  MAIAU              &  62.44 &    16 &   11300 &   1.68 &  B9             &   377443211 &  MAIAU              &  61.19 &     1 &   12600 &   1.52 &  B8V            \\
 137822799 &  MAIAH              &  73.99 &     1 &   10265 &   2.20 &  B9.5III        &   379937109 &  MAIAU              &  50.78 &    35 &   11300 &   1.39 &  B9             \\
 144517863 &  MAIAU              &  52.95 &    17 &   12151 &   2.20 &  B9V            &   387757610 &  MAIAH              &  54.86 &     7 &   10422 &   1.53 &  A2             \\
 144710346 &  MAIAU              &  55.58 &     4 &   10460 &   1.16 &  A3             &   388688820 &  MAIAU              &  50.15 &    14 &   10089 &   1.26 &  A3             \\
 144956901 &  MAIAH              &  57.33 &     2 &   13700 &   3.02 &  B7III          &   391154611 &  MAIAH              &  51.52 &     4 &   10300 &   1.54 &  B9V            \\
 145923579 &  MAIAU              &  62.88 &     2 &   10435 &   1.66 &  B9IV-Vkn       &   391346342 &  SPB+MAIAU          &  59.59 &     3 &   13932 &   2.74 &  B6Vnn          \\
 174662768 &  SPB+MAIAH          &  50.64 &     3 &   14062 &   2.71 &  B5Vn           &   401536404 &  MAIAH              &  53.17 &     1 &   10298 &   2.16 &  B9.5Vn         \\
 174866532 &  MAIAH              &  51.87 &     1 &   10288 &   1.19 &  A2             &   401785909 &  MAIAU+ROT          &  52.66 &     5 &   10638 &   1.24 &  A2             \\
 182910557 &  MAIAU              &  60.43 &     1 &   10063 &   1.20 &  A3/5           &   403787892 &  MAIAH+EA           &  57.32 &     7 &   16143 &   3.30 &  B5III          \\
 183522571 &  MAIAU              &  60.16 &     5 &   10799 &   1.45 &  A1/2IV/V       &   415545142 &  MAIAH              &  55.56 &     2 &   10184 &   1.29 &  A2             \\
 196500681 &  MAIAH              &  52.07 &     1 &   11100 &   1.62 &  B9.5V          &   416635930 &  MAIAH+EB           &  51.03 &     6 &   12331 &   2.55 &  B9IV-V         \\
 202121249 &  MAIAH              &  51.49 &     4 &   11300 &   1.30 &  B9:V           &   419367293 &  MAIAH              &  51.98 &     1 &   11300 &   2.24 &  B9V            \\
 202431888 &  MAIAU              &  74.11 &     1 &   12023 &   2.53 &  B9IVSi:        &   421714420 &  MAIAU+EA           &  53.15 &     6 &   12600 &   2.33 &  B8             \\
 220313579 &  MAIAU              &  61.50 &     9 &   10014 &   1.35 &  A2/3           &   422454176 &  MAIAH+ROT          &  50.46 &     2 &   12600 &   1.51 &  B8.5V          \\
 233110625 &  MAIAH              &  52.56 &     1 &   10199 &   1.57 &  A2             &   429306233 &  MAIAU              &  54.76 &     5 &   10000 &   1.79 &  A0p(Si)        \\
 233310793 &  MAIAH              &  51.90 &    21 &   12678 &   1.55 &  A5-F5          &   440638544 &  MAIAU              &  58.58 &     6 &   10561 &   1.18 &  A0             \\
 236785664 &  MAIAU+BE+ROT       &  56.97 &     4 &   11300 &   2.45 &  B9Ve           &   447086833 &  MAIAH              &  50.38 &     1 &   11300 &   1.45 &  B9/A1V         \\
 238317411 &  MAIAU+ROT          &  69.52 &     1 &   11300 &   2.20 &  B9IV           &   450784713 &  MAIAU+ROT          &  58.43 &    14 &   10128 &   1.79 &  B9/A0          \\
 238641255 &  MAIAU+ROT+FLARE    &  61.82 &    25 &   10137 &   1.21 &  A1V            &   464470984 &  MAIAH+ROT          &  59.70 &     1 &   11066 &   2.17 &  B9IV/V         \\
 251058870 &  MAIAH              &  54.75 &     4 &   11300 &   2.26 &  B9III:nn       &   469421586 &  MAIAH+BE+ROT       &  53.86 &     1 &   12600 &   2.91 &  B8IV/V         \\
\hline                        
\end{tabular}
}
\end{center}
\end{table*}

\begin{figure}
\begin{center}
\includegraphics{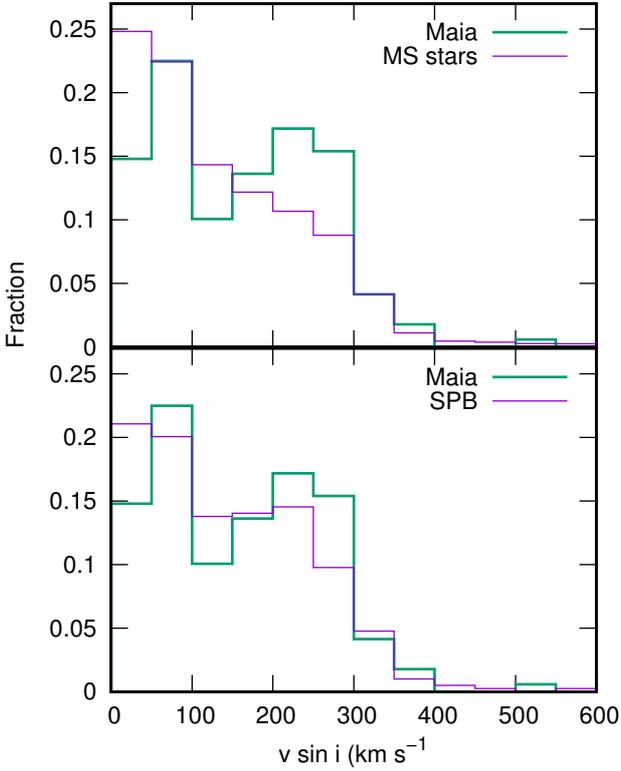}
\caption{Top panel: the distribution of projected rotational velocity, $v\sin i$, for
Maia variables (heavy green histogram) and for main sequence stars in the same 
temperature range. The bottom panel shows the distributions for Maia and SPB
stars in the same temperature range.} 
\label{vsini}
\end{center}
\end{figure}

\begin{figure}
\begin{center}
\includegraphics{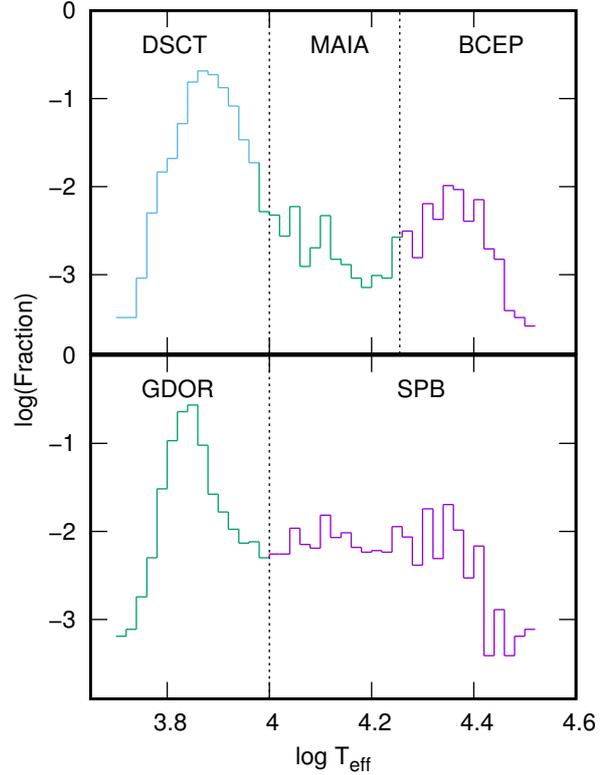}
\caption{The distribution of $\delta$~Sct, Maia and $\beta$~Cep variables
(top panel) and the $\gamma$~Dor and SPB stars (bottom panel).}
\label{vden}
\end{center}
\end{figure}

\section{Rotation}

As already mentioned, rotation affects the apparent location of the star in
the H--R diagram due to equatorial gravity darkening and introduces new 
gravito-inertial modes with moderately high frequencies \citep{Salmon2014}.
Gravity darkening will occur in all stars, whether pulsating or not.
Therefore no bias is introduced in comparing the rotational velocities of
Maia stars with main sequence stars in the same effective temperature range.

The distribution of projected rotational velocities, $v\sin i$, for 169 Maia 
stars and for 6546 main sequence stars in the temperature range ($10000
< T_{\rm eff} < 18000$) is shown in Fig.\,\ref{vsini}.  Also shown in the
same figure is a comparison with 399 SPB stars in the same temperature range.
The conclusion is clear: Maia stars are not rapidly rotating SPB stars.  The
rotation rates of both Maia and SPB stars are the same as in normal main
sequence stars.

\section{Distribution of Maia stars}

Fig.\,\ref{vden} shows the number density of $\delta$~Sct, Maia and $\beta$~Cep
as well as $\gamma$~Dor and SPB variables as a function of effective 
temperature. This figure was constructed by counting the number of stars of a 
particular variability class within a small range of effective temperature 
(0.02\,dex in $\log T_{\rm eff}$)  within the main sequence band as
previously defined.

The figure illustrates the problem pointed out by \citet{Balona2020a}: there 
are no distinct instability regions among hot stars. There is no doubt 
that uncertainty in the effective temperature must be partly responsible.
The typical uncertainty in $\log T_{\rm eff}$ is about 0.02\,dex 
(which is the bin size in the figure), so it seems unlikely to be the sole 
reason. The figure also appears to suggest that Maia stars are not a separate 
class, but simply an extension of the $\delta$~Sct variables to mid-B spectral 
types. On this basis one could rename Maia variables as ``hot $\delta$~Sct 
stars'' or something similar.  However, this would be contrary to the spirit 
of classification in which stars of a particular group pulsate with the same
mechanism.

Fig.\,\ref{vden} shows that the $\gamma$~Dor and SPB variables form a continuous
group, quite unlike predictions from stellar models. Moreover, it has been
established that the $\gamma$~Dor instability region lies within the  
$\delta$~Sct instability region \citep{Balona2018c}, further eroding
the idea that different stellar variability classes are associated with 
different pulsation mechanisms. 

As already mentioned, \citet{Sharma2022} suggested that gravity darkening due to 
rapid rotation might explain B-type low-frequency pulsators 
too cool to be SPB stars.  To test this idea, we examined the $v\sin i$ 
distribution of 60 low-frequency pulsators with $9000 < T_{\rm eff} < 11000$\,K
and compared it to the $v\sin i$ distribution of 1217 main sequence stars in 
the same effective temperature range.  There is no obvious difference between 
the distributions, but the sample is rather small.  For SPB stars, the mean
is $\langle v\sin i\rangle = 159 \pm 12$\,km\,s$^{-1}$ and for main 
sequence stars $\langle v\sin i\rangle = 118 \pm 2$\,km\,s$^{-1}$.  In any case
the rotation rate of these hot $\gamma$~Dor/SPB variables cannot be called
rapid.

Pulsations in $\delta$~Sct stars are driven by the opacity $\kappa$~mechanism 
operating in the He\,II ionization zone.  This mechanism does not work for 
effective temperatures as high as $T_{\rm eff} \approx 16000$\,K, which, 
judging from Fig.\,\ref{vden}, is the highest temperature for pulsation in 
Maia stars.  According to the latest models, pulsational driving of low degree 
modes in $\delta$~Sct stars does not occur for $T_{\rm eff} > 9000$\,K 
\citep{Xiong2016}.  According to current models, Maia stars cannot be
considered as high temperature $\delta$~SCt stars, nor can they be explained
as rapidly rotating SPB stars.

\section{Conclusions}

From {\em TESS} light curves, 493 stars have been identified as Maia
variables. These are defined as main sequence stars with $10000 < T_{\rm
eff} < 18000$\,K and frequencies greater than 5\,d$^{-1}$.  By comparing the
projected rotational velocities of the stars with those of main sequence stars
or SPB variables in the same temperature range, it is shown that Maia stars are
not rapidly rotating.  Therefore the idea that the high frequencies in Maia
variables are due to rapid rotation \citep{Salmon2014} is refuted.  Moreover,
such an idea cannot explain frequencies of over 50\,d$^{-1}$ seen in many Maia
stars.

It is also demonstrated that there is no clear separation in effective
temperature between $\delta$~Sct and Maia variables or between $\beta$~Cep
and Maia variables.  Indeed, as Fig.\,\ref{vden} shows, there is no clear
separation between any variability group among early-type stars, as already
pointed out in \citet{Balona2020a}.  The sequence of SPB stars, which begins
at about spectral type B0,  runs all the way to A0 where it merges with the
sequence of $\gamma$~Dor stars running towards the cool end of the
$\delta$~Sct instability strip (the hot $\gamma$~Dor variables, 
\citealt{Balona2016c}).  The largest concentration of $\gamma$~Dor
stars occurs around $T_{\rm eff}\approx 7000$\,K.  The instability strip defined
by the majority of $\gamma$~Dor stars actually lies within the $\delta$~Sct
instability strip, as pointed out by \citet{Balona2018c}, which is very
puzzling. 

\citet{Sharma2022} suggested that low frequency stars between the SPB and 
$\delta$~Sct instability strips are probably SPB stars shifted to cooler 
temperatures owing to gravitational darkening at the equator as a result of 
rapid rotation.  This not supported by a comparison of their projected 
rotational velocities with those of main sequence stars in the same effective 
temperature range.  These low frequency late-B stars merge into the hot
$\gamma$~Dor variables.  Rapid rotation certainly cannot be an explanation for 
the hot $\gamma$~Dor stars, since the required apparent temperature shift
due to equatorial gravity darkening is far too large.

While uncertainties in $T_{\rm eff}$, different metallicities and other
factors may be partly responsible for the lack of well-defined instability 
regions, it seems unlikely that these are the sole causes. There is clearly 
something wrong with current pulsation models.  Since the models cannot 
explain the easily identifiable $\gamma$~Dor, $\delta$~Sct, SPB, Maia and 
$\beta$~Cep stars, it is difficult to place much trust on models involving 
deep internal core pulsations (e.g. \citealt{Lee2020b}).

It is possible that there may be an interplay of different driving mechanisms 
which vary in effectiveness. The discovery that starspots are pervasive along 
the entire main sequence \citep{Balona2022b} is relevant. This seems to imply
that surface convection present in all A and B stars. Including convection
in the models may assist in resolving these problems, but convection may not 
be the only factor. 

A better understanding of the upper stellar layers of A and B stars is
required in order to improve the pulsation models. Therefore, an in-depth study 
of the causes of rotational light modulation is required.  In order to 
understand the roAp-like frequencies in some Maia stars (Table\,\ref{maiau}), 
modes of high degree should receive particular attention.

\section*{Acknowledgments}

LAB wishes to thank the National Research Foundation of South Africa for 
financial support and Dr Tim Bedding for useful comments.

Funding for the {\it TESS} mission is provided by the NASA 
Explorer Program. Funding for the {\it TESS} Asteroseismic Science Operations 
Centre is provided by the Danish National Research Foundation (Grant agreement 
no.: DNRF106), ESA PRODEX (PEA 4000119301) and Stellar Astrophysics Centre 
(SAC) at Aarhus University. 

This work has made use of data from the European Space Agency (ESA) mission 
Gaia, processed by the Gaia Data Processing and Analysis Consortium (DPAC).
Funding for the DPAC has been provided by national institutions, in particular 
the institutions participating in the Gaia Multilateral Agreement.  

This research has made use of the SIMBAD database, operated at CDS, 
Strasbourg, France.  Data were obtained from the Mikulski Archive for Space 
Telescopes (MAST).  STScI is operated by the Association of Universities for 
Research in Astronomy, Inc., under NASA contract NAS5-2655.

\section*{Data availability}

All data are incorporated into the article is available though the author and 
described in \citet{Balona2022c}.

\bibliographystyle{mnras}
\bibliography{maiatess}

\label{lastpage}

\end{document}